# On the Heisenberg principle at macroscopic scales: understanding classical negative information. Towards a general physical theory of information

*Álvaro López-Medrano*

Abstract

With the aid of a toy model, the Monty Hall Problem (MHP), the counterintuitive and theoretically problematic concept of negative information in classical systems is well understood. It is shown that, as its quantum counterpart, classical local mutual information, obtained through a measurement, can be expressed as the difference between the information gained with the evidence and the negative information generated due to the inefficiency of the measurement itself; a novel local Shannon´s metric, the *transfer information content*, is defined as this difference, which is negative if the measurement generates more disturbance than the evidence, i.e., generates a classical measurement back action. This metric is valid for both, Classical and Quantum measurements, and it is proposed as a starting point towards a general physical theory of information. This information-disturbance trade-off in classical measurements is a kind of Heisenberg principle at macroscopic scales, and it is proposed, as further work, to incorporate this result in the already existing generalized uncertainty principles in the field of quantum gravity.

## 1. INTRODUCTION

In (Wiseman, 2009), it is stated that *"measurement and probability do not play a significant role in the foundation of classical mechanics (although they do play a very significant role in practical applications of classical mechanics, where noise is inevitable)"*. It is shown that, even in noiseless schemes, measurement and probability are in the essence of classical systems.

Paradoxically, it is stated as well in (Wiseman, 2009) that Quantum measurement theory *"provides the essential link between the quantum formalism and the familiar classical world of macroscopic apparatuses"*.

With the aid of a toy model (the Monty Hall Problem), and bringing Shannon´s Information Theory to statistics, it is shown that the information-disturbance trade-off described for quantum measurements (Fuchs, 1996) (Dariano, 2003) (Funo, 2013) (Terashima, 2016), also takes place in classical measurements, and that, in both cases, leads to negative informative measurements. Therefore, this concept of negative information, which is considered as counterintuitive and theoretically problematic (Evans, 1996), confers classical systems the weirdness of quantum ones, and thus links both.

Based on this link between classical and quantum systems, a Generalized Theory of Information, and the existence of a Heisenberg principle at macroscopic scales, are postulated in this work.

The work is organized as follows: in Section 2 the statistical meaning of Shannon´s Information Measures is explained; in Section 3 the concept of Shannon´s information content is described; in Section 4 Bayes´ Theorem is expressed in terms of its information contents; Section 5 gives an interpretation of the mentioned Bayes´ Theorem information contents and introduces the Shannon´s metric called *transfer information content*; in Section 6, well known Shannon´s metrics, like the Kullback-Leibler divergence and the mutual information, are expressed in terms of the *transfer information content*; in Section 7, a toy model, the Monty Hall Problem, is solved in terms of the *transfer information content* and the concept of negative local mutual information is understood; in Section 8, the quantum measurement is expressed in terms of the transfer information content; in Section 9, the quantum indistinguishability is explained based on the transfer information content, and it is shown how the quantum back action generates negative unaveraged mutual information; Section 10 proposes formulas for a Generalized Theory of Information; in Section 11, several aspects of the work are discussed; Section 12 outlines the conclusions.

## 2. STATISTICAL MEANING OF SHANNON´S INFORMATION MEASURES

Shannon´s Information Theory publication (Shannon, 1948) attracted the attention of many statisticians and physicists (Soofi, 2000) like Kullback, Lindley, Zellner, Jaynes and Akaike. The reason of their interest on Shannon´s work was the fact that the mentioned theory defined information in terms of probability, and they applied Shannon´s information concepts to classic probability theorems like Bayes. As a result of this statistical approach they derived the so called Shannon´s information functions like the Kullback-Leibler divergence (Kullback, 1951) or the Zellner´s information processing rule (Zellner, 2002) between others, see (Nelson, 2008) for an excellent review of Shannon´s information metrics.

However, the information meaning that was derived by statisticians and physicists was far from the one proposed by Shannon, with great success, for communications engineering (Massaro, 1993). In fact, it has become clear that the statistical meaning of information measures a facet of information that is different and complementary to Shannon's definition (Baldi, 2010), and this is why this generalization of Shannon´s Information Theory in statistics has attracted the attention of information processing psychologists (Massaro, 1993), cognitive scientists (Nelson, 2008), philosophers (Bandyopadhyay, 2010), decision making economists (DeBondt, 1995) and quantum physicists who have coined the term Quantum Information Theory (Horodecki, 2003).

The work presented in this paper is again an example of bringing Shannon´s Information Theory to statistics in order to capture aspects of information that the mentioned theory does not capture. In particular the present research exploits, applying it in Bayes´ Theorem, a concept derived from Shannon´s Information Theory and defined in (Hehner, 2011): the information content of each message (or each event) individually (we will clarify this concept in detail later in the paper); Shannon himself was reluctant to talk about the mentioned concept (Hehner, 2011), but there is no harm in doing so (Hehner, 1977). The interpretation of the information content of the terms that appear in Bayes´ Formula is the main contribution of this research, and it will explain how the information provided by an experiment is processed, as well as, the concept of negative information which is considered as counterintuitive and theoretically problematic (Evans, 1996), in fact one issue in the literature is whether it is better to have a measure of information that is always positive (nonnegativity) (Nelson, 2008), this work shows that there should be nothing against negative information metrics.

The Monty Hall Problem (MHP) (Rosenhouse, 2010), in some of its different versions, is used in this work as an excellent example to show the correctness of the interpretations of the mentioned information contents in Bayes´ Formula. The resolution of the mentioned MHP presented here shows, as well, how problems that have usually been addressed in the probability framework can be solved in the information one (Hehner, 2011).

## 3. INFORMATION CONTENT CONCEPT REVIEW

The main contribution of the mentioned Information Theory is the Shannon´s Entropy, also known as self-entropy, of a discrete random variable X *(in this work only discrete random variables and discrete Shannon´ Functions are considered, however the results and concepts derived are applicable to the continuous equivalents)*. The entropy is measured in bits and defined as (Shannon, 1948):

$$H(X) = \sum_{i=1}^{n} -p(x_i) \cdot log_2[p(x_i)] \qquad (1)$$

The entropy in (1) is often called the average total information (Shannon, 1948), so that,

$$H(X) = \bar{I}_T \qquad (2)$$

As pointed out above, Shannon was reluctant to talk about the information content of each message (or each event) individually (Hehner, 2011), but there is no harm in doing so (Hehner, 1977), in (Hehner, 2011) the information content (measured in bits) of the event $x_i$, also known as self-information, and is given by:

$$I(x_i) = -log_2[p(x_i)] bits \qquad (3)$$

And then equation (1) can be expressed in terms of the information contents of each event as:

$$H(X) = \sum_{i=1}^{n} p(x_i) \cdot I(x_i) \qquad (4)$$

So, it is equivalent to talk about the probability of the event $x_i$, than to talk about its information content (Hehner, 2011), in fact we can express the probability in terms of the information content, from (3):

$$p(x_i) = 2^{-I(x_i)} \quad (5)$$

Shannon explained the amount of information carried by a message as a measure of how surprised one is to learn the message (Shannon, 1948). Thus, it is natural to apply the same concept to the *unaveraged* information content given above in equation (3), i.e., it can be stated that this metric represents, in bits, the surprise that provokes the occurrence of a given event. From equations (3) and (5) it is straight forward that the higher/lower the probability of a given event the lower/higher its information content, i.e., the surprise it provokes. This concept of surprise will be used here although not in the same sense as the one proposed in (Baldi, 2010) to which we will refer later in the paper.

As pointed out again in (Hehner, 2011) nowadays it makes more sense to talk in terms of information than in terms of probability, most people today have a quantitative idea of what information and memory are. This paper confirms this assertion showing that by expressing classic probability theorems in terms of Shannon´s information measures, it is possible to understand the way information is transferred, and how classical negative information is generated.

## 4. BAYES´ THEOREM IN TERMS OF ITS INFORMATION CONTENTS

Consider the Bayesian approach to data modeling and inference, well described in (Lindley, 1956); Lindley was the first to develop a measure of information in data $x$ belonging to the space $X$ about a parameter $\theta$ that ranges over the current space of hypothesis or models $\Theta$ with the prior distribution $f(\theta)$ (Soofi, 2000). Lindley adopted Shannon´s mutual information for measuring the expected information in data $x$ about $\theta$ (Soofi, 2000), and enunciated several Theorems related to the concept of the information gain due to data $x$, belonging to a space $X$ that are of interest for this work; the notation in Lindley´s paper will be maintained in this one.

The effect of data $x$ on the observer is to change the prior distribution $P(\theta)$, into the posterior distribution $P(\theta|x)$ via Bayes´ Theorem:

$$P(\theta|x) = P(x|\theta)P(\theta)/P(x) \quad (6)$$

where $P(x)$ is known as the evidence, and $P(x|\theta)$ as the likelihood of Bayes´ formula. And then, if we take the $-log_2$ function of both sides in equation (6) we have:

$$-log_2[P(\theta|x)] = -log_2[P(x|\theta)P(\theta)/P(x)] \quad (7)$$

And then applying the properties of the $log$ function we can expand the right-hand side of equation (7) and we have:

$$-log_2[P(\theta|x)] = \\ = -log_2 P(x|\theta) - log_2 P(\theta) + log_2 P(x) \quad (8)$$

And taking into account equation (3) we can express equation (8) in terms of the information contents corresponding to each of the probabilities that appear in Bayes´ Formula (note that the order of the terms in right hand side of equation (8) have been changed for reasons that we will explained below):

$$I(\theta|x)\ (bits) = I(\theta)(bits) - I(x)(bits) + I(x|\theta)(bits) \quad (9)$$

Remind that, in this last equation (9), each of the terms is the information content of a certain event and it represents, in bits, the surprise that provokes the occurrence of the mentioned event. The interpretation of this last equation (9) is the main contribution of this paper and, as stated before, it will allow to understand how the bits of information, introduced in a space of models, are transferred to each of them, and concepts like negative information, often not well understood (Evans, 1996), will arise very naturally.

## 5. INTERPRETATION OF INFORMATION CONTENTS IN BAYES´ THEOREM. TRANSFER INFORMATION CONTENT METRIC

From equation (9) it can deduced that the effect of data on the observer is to change the information content of a given event $\theta$ in the space of models, i.e., it changes (increases or decreases) the surprise that provokes the occurrence of the mentioned event. Therefore, we call the term on the left hand side in equation (9), $I[P(\theta|x)]$, the *Posterior Information Content* of a single model $\theta$, and then the first term of the right hand side in that equation, $I[P(\theta)]$, is called the *Prior Information Content of a single model* $\theta$.

The key aspect about equation (9) is that the mentioned change of information content, of a given event in the space of models, that provokes a *single* observation $x$, is not compounded, as our intuition tells us, of a single information content, but by two information contents $I[P(x)]$ and $I[P(x|\theta)]$ which are generated by the mentioned single observation; once we give an interpretation of each of these two components in equation (9), we will give an interpretation of the previous mentioned fact that *from a single observation, not one, but two information contents appear*.

Based on equation (3) it can be deduced that information contents are always positive and thus, taking into account the signs in equation (9), $-I[P(x)]$ is always negative, i.e., it can only contribute (if different from zero) to reduce *the Posterior Information Content* $I[P(\theta|x)]$, i.e., it can only contribute to increase the probability of $\theta$ once $x$ is observed. Note that, although this term is always negative, it adds information about $\theta$ since it contributes to increase its probability of occurrence, that is why we call this term $-I[P(x)]$ the *Positive Information Content Associated to the Evidence* $P(x)$.

The opposite can be said about $+I(x|\theta)$: it is always positive or zero and then, it can only contribute (if different from zero) to increase the *Posterior Information Content* $I(\theta|x)$, i.e., it can only contribute to decrease the probability of $\theta$ once $x$ is observed. Reasoning in the same way as in the previous paragraph, it can be stated that this term subtracts information about $\theta$ since it contributes to decrease its probability of occurrence, and that is why we call this term $+I(x|\theta)$ the *Negative Information Content Associated to the Bayes´ Likelihood* $P(x|\theta)$.

Now, once an interpretation for the two last terms in equation (9) has been given, we can come back to the fact that a single observation provides two information contents that affect the Posterior Information Content $I(\theta|x)$: what equation (9) is telling us is that every observation carries a positive information associated to the evidence $P(x)$ throughout the term $-I(x)$, but that at the same time there is a *cost* for carrying that positive information which produces a negative information associated to the likelihood $P(x|\theta)$ throughout the term $+I(x|\theta)$.

Therefore since $-I(x)$ is negative information, and $+I(x|\theta)$ is positive information, the net result can be positive or negative depending on the absolute value of each of the two terms and we define the metric called *transfer information content from $x$ to $\theta$* as the difference between these two information contents (in the formulas we will refer to the *Transfer Information Content* by its acronym TIC):

$$TIC(x \rightarrow \theta)(bits) = I(x) - I(x|\theta) \qquad (10)$$

We have the following three cases:

1. $I(x) > I(x|\theta) \Rightarrow TIC(x \rightarrow \theta) > 0$: the effect of data $x$ on the observer is to decrease the a posteriori information content of the event $\theta$, i.e., to increase its probability
2. $I(x) = I(x|\theta) \Rightarrow TIC(x \rightarrow \theta) = 0$: there is no effect of data on the observer.
3. $I(x) < I(x|\theta) \Rightarrow TIC(x \rightarrow \theta) < 0$: the effect of data $x$ on the observer is to increase a posteriori information content of the event $\theta$, i.e., to decrease its probability

As a summary, we can conclude that the evidence is informative depending on the likelihood in Bayes´ Formula. As shown below, this is confirmed by the well-known metric *Bayes Factor*.

The *Bayes Factor* (Kass, 1995) is the ratio of the likelihood, based on observed data $x$, associated to two different models $\theta_1$ and $\theta_2$:

$$K = \frac{P(x|\theta_1)}{P(x|\theta_2)} \qquad (11)$$

And the log version of this last equation is,

$$\log_K = log_2 \left[\frac{P(x|\theta_1)}{P(x|\theta_2)}\right] \qquad (12)$$

The *Bayes Factor*, as can be deduced from equation (11), can be interpreted as a measure of the *weighting of evidence* (Good, 1950). We highlight this metric because it confirms what has been proved before: that the *effectiveness of the evidence* depends on the likelihood correspondent to a given model since the later always provided negative information, so that the lower/ higher it is, the more or less *informative* is the observation for the model, precisely as the Bayes Factor confirms.

# 6. KL DIVERGENCE AND MUTUAL INFORMATION IN TERMS OF THE TRANSFER INFORMATION CONTENT

The well-known Shannon´s metric Kullback-Leibler divergence (Kullback 1951) measures the average information gain of a single observation $x$ over the space of models $\Theta$, and it is given by the following expression:

$$D_{KL}\{[(\Theta|X)\|\Theta]\} = \sum_\theta P(\theta|x) log_2 \left[\frac{P(\theta|x)}{P(\theta)}\right] \qquad (13)$$

From Bayes´ Theorem, or equivalently from equation (9), we have this analog expression:

$$D_{KL}\{[(\Theta|X)\|\Theta]\} = \sum_\theta P(\theta|x) log_2 \left[\frac{P(x|\theta)}{P(x)}\right] \qquad (14)$$

And expanding the logarithm:

$$D_{KL}\{[(\Theta|X)\|\Theta]\} = \\ = \sum_\theta P(\theta|x) \cdot \{log_2[P(x|\theta)] - log_2[P(x)]\} \qquad (15)$$

And, considering equation (3), this last expression can be expressed in terms of the information contents associated to the probabilities involved:

$$D_{KL}\{[(\Theta|X)\|\Theta]\} = \sum_\theta P(\theta|x)\cdot\{I(x) - I(x|\theta)\} \quad (16)$$

And from equation (10):

$$D_{KL}\{[(\Theta|X)\|\Theta]\} = \sum_\theta P(\theta|x)\cdot\ TIC(x\to\theta) \quad (17)$$

This last two equations (16) and (17) express the Kullback-Leibler divergence as the *average transfer information content* of a single observation $x$ over the space of models $\Theta$. This interpretation of the mentioned KL divergence coincides with the one given in (Schreiber, 2000), where this metric is coined as *transfer entropy,* defined in terms of the *transfer information content* in equations (16) and (17).

If in equation (17) we average (integrate) over the space of data $X$, we obtain the *Shannon´s Mutual Information* in terms of the *transfer information content*:

$$I(\Theta, X) = H(\Theta) - H(\Theta|X) = E_x\{D_{KL}[(\Theta|X)\|\Theta]\} = E_x\{\sum_\theta P(\theta|x)\cdot\ TIC(x\to\theta)\} \quad (18)$$

This last equation (18) says that the *mutual information* is the average over the space of data of the *average transfer information content* over the space of models.

Now, if in equation (13) we consider a single model $\theta$ instead of the average over all of them, we have:

$$log_2\left[\frac{P(\theta|x)}{P(\theta)}\right] = I[P(x)] - I[P(x|\theta)] = TIC(x\to\theta) \quad (19)$$

The log-odd ratio on left hand side of Equation (19), $log_2\left[\frac{P(\theta|x)}{P(\theta)}\right]$, has been recently coined, inside the emergent field of *Local Information Dynamics* (Lizier, 2010), as the *local transfer entropy* in (Lizier, 2008). Although it coincides numerically with the *transfer information content*, its expression, given by the mentioned log-odd ratio, does not allow to derive the interpretation of how information is transferred given in Section 4, and the reason for this is that, although it indeed represents a transfer of information, it is expressed in terms of probabilities, i.e., in the probability framework and not in the information one, as it is the mentioned *transfer information content* defined in this work.

Again, the same the log-odd ratio in Equation (19) is called in (Baldi, 2010) the *single model surprise* and it is measured in *wows* instead of in bits. What is stated in this work is that mentioned single model surprise is equivalent to the *transfer information content*, and then, it can be measured in bits.

# 7. MONTY HALL PROBLEM SOLVED BASED ON THE TRANSFER INFORMATION CONTENT (UNDERSTANDING NEGATIVE CLASSICAL INFORMATION)

The previous interpretation of the information contents in equation (9) is going to be clarified with the MHP, this is the formulation in its traditional form (Rosenhouse, 2009):

*Suppose you are a contestant on a quiz show. The host, Monty Hall, shows you the three doors (A, B, and C). Behind one door is an expensive new car and behind others are goats. You are to choose one door. If you choose the door with the car, you get it as a prize. If you choose a door with a goat, you get nothing. You announce your choice, and Monty Hall opens one of the other two doors, showing you a goat, and offers to let you change your choice. Should you change? Three crucial points need to be clearly stated which are usually overlooked in a popularized version of the MHP (Bandyopadhyay, 2010). They are (i) the expensive car has an equal chance of being distributed behind any of the doors. (ii) If you choose one door without the prize behind it, Monty will open the door that does not have the prize behind it. And (iii) if you choose the door in which there is a prize behind it, then Monty will open the door randomly which does not have the prize behind it. Suppose you have chosen door A. Would you switch or stay? Solution: if Monty opens door B: P(A)=1/3, P(C)=2/3; if Monty opens door C: P(A)=1/3, P(B)=2/3 (Rosenhouse, 2009).*

In the previous reference (Rosenhouse, 2009) it is envisaged to apply Shannon´s Information Theory to the MHP. As it will be shown below, the concepts explained in the previous Sections of this work and derived from the mentioned Theory, find an excellent environment in the MHP).

Firstly, we are going to compute the *Posterior Information Content* of door A using equation (9). The following probabilities are needed: $P(x|\theta), P(x)$, where the observation $x$ takes place when Monty opens one of the two doors B or C (without loss of generality we assume that Monty opens door B), and $\theta$ is the event correspondent to opening a door and finding a car behind it.

Now, from the formulation of the problem, we have three models A, B and C with the following prior distribution,

$P(\theta = A) = \frac{1}{3}; P(\theta = B) = \frac{1}{3}; P(\theta = C) = 1/3$ (20)

$P(x = MontyB) = P(Monty_B|A) \cdot P(A) + P(Monty_B|B) \cdot P(B) + P(Monty_B|C) \cdot P(C) = \frac{1}{2} \cdot \frac{1}{3} + 0 + 1 \cdot \frac{1}{3} = \frac{1}{2}$ (21)

$P(Monty_B|A) = \frac{1}{2}$ (22)

And then applying equation (9),

$I(\theta = A|x = MontyB)\ (bits) = I(A)(bits) - I(MontyB)(bits) + I(MontyB|A)(bits) = -log_2\left(\frac{1}{3}\right) + log_2\left(\frac{1}{2}\right) - log_2\left(\frac{1}{2}\right) = -\log\left(\frac{1}{3}\right) bits$ (23)

And now computing the probability associated to $I(A|MontyB)$, applying equation (5),

$P(A|MontyB) = 2^{-[I(A|MontyB)]} = 1/3$ (24)

Now, if we wanted to compute the *Posterior Information Content* of door C, again using equation (9), we have:

$P(Monty_B|C) = 1$ (25)

$I(\theta = C|x = MontyB)\ (bits) = I(C)(bits) - I(MontyB)(bits) + I(MontyB|C)(bits) = -log_2\left(\frac{1}{3}\right) + log_2\left(\frac{1}{2}\right) - log_2(1) = -\log\left(\frac{2}{3}\right) bits$ (26)

And then $P(C|MontyB)$ is given by:

$P(C|MontyB) = 2^{-[I(C|MontyB)]} = 2/3$ (27)

Note that the MHP solution that we have derived in equations (24) and (27) coincides with the well-known counterintuitive solution (Rosenhouse, 2009). To solve the MHP we have not made use of the Bayes´ Theorem in its probability form but in its information content form given in equation (9); as we are going to see below, this new version of Bayes´ Theorem allows to suppress the counterintuitive character usually associated to the results that the mentioned Theorem produces (Rosenhouse, 2009) and to understand the concept of negative information in classical systems.

Now let us compare what happens, in door A and C, in terms of information transfer when Monty opens door B: the information content associated to the evidence, i.e., the probability of Monty opening door B is the same in both cases ($-log_2\left(\frac{1}{2}\right) = 1bit$); however in the case of door A this information is not transferred to it, i.e., does not contribute to reduce its information content, i.e., to increase the probability of finding the car behind this door; the reason for this is, as explained in the previous section, the negative information content associated to the likelihood, $+I(MontyB|A) = 1bit$. So in this case the positive information content generated by the evidence is cancelled by the likelihood, we could say that the contestant is paying the cost of Monty´s opening randomly door B and *it is not counterintuitive that this Monty´s randomness is introducing negative information in the system*. Thus, the *transfer information content* of door A due to Monty opening B, as expected, is null:

$TIC(MontyB \rightarrow A) = I(MontyB)(bits) - I(MontyB|A)(bits) = 1 - 1 = 0bits$ (28)

On the other hand, in the case of door C the information content associated to the evidence, 1bit, is not cancelled at all because the likelihood, $P(MontyB|C)$, is equal to 1, and then its information content is null, i.e., Monty does not introduce any negative informative bits, and the positive information of the evidence is completely transferred to this door decreasing its information content and increasing its associated probability. And, therefore, the *transfer information content* of door C is 1bit and then equal to the information content of the evidence, i.e., the later has been totally informative for door C:

$TIC(MontyB \rightarrow C) = I(MontyB)(bits) - I(MontyB|C)(bits) = 1 - 0 = 1bit$ (29)

Now we can compute the KL Divergence as the *expected transfer information content*, applying equation (16) we have,

$D_{KL}\{[(\Theta|X = MontyB)\|\Theta]\} = \sum_\theta P(\theta|MontyB) \cdot TIC(MontyB \rightarrow \theta) = P(A|MontyB) \cdot TIC(MontyB \rightarrow A) + P(C|MontyB) \cdot TIC(MontyB \rightarrow C) = \frac{1}{3} \cdot 0 + \frac{2}{3} \cdot 1 = 0,\hat{6}bits$ (30)

In order to compute the mutual information based on equation (18) we take into account that the space of data X is equiprobable and then the *average transfer information contents* (or the KL divergence) generated by Monty, when opening doors B or C, are similar and equal to $0,6bits$. So we have,

$I(\Theta, X) = E_x\{\sum_\theta P(\theta|x) \cdot\ TIC(x \rightarrow \theta)\} = P(MontyB) \cdot 0,\hat{6} + P(MontyC) \cdot 0,\hat{6} = \frac{1}{2} \cdot 0,\hat{6} + \frac{1}{2} \cdot 0,\hat{6} = 0,\hat{6}bits$ (31)

It can be shown that the results in equations (30) and (31) coincide with the ones that would have been obtained with the classical expressions of the *KL divergence* and the *mutual information* respectively.

Now consider a variant of the traditional MHP, this is its formulation:

*Biased MHP: Monty rolls a 6-sided die, and if it comes up 1, 2, or 3, he hides the car behind door A; if it comes up 4 or 5 he hides the car behind door B, and if it comes up 6 he hides the car behind door C. And the contestant knows that this is Monty's method of hiding the car. So, the contestant naturally chooses door A. Now Monty acts as before: if the car is behind door A, Monty opens either of door B or C at random; if the car is behind either door B or C, Monty opens the other one. Should the contestant stick or switch? Answer: if Monty opens door B: P(A)=3/5, P(C)=2/5, the contestant should not switch; if Monty opens door C: P(A)=3/7, P(B)=4/7, the contestant should switch. (Rosenhouse, 2009)*

This MHP variant is a very interesting for our purposes, because it helps to understand the concept of negative information in classical systems.

We again assume that the contestant chooses the door A, which in this case is reasonable, since this has the maximum prior probability. In this case, the solution differs depending on Monty opening door B or C. We are going to consider the case in which Monty opens door C since it gives the most interesting results.

Now, proceeding as in the traditional case we have,

$P(\theta = A) = \frac{1}{2}; P(\theta = B) = \frac{1}{3}; P(\theta = C) = 1/6$

$P(Monty_C) = P(Monty_C|A) \cdot P(A) + P(Monty_C|B) \cdot P(B) + P(Monty_C|C) \cdot P(C) = \frac{1}{2} \cdot \frac{1}{2} + 1 \cdot \frac{1}{3} + 0 = \frac{7}{12}$ (32)

$P(Monty_C|A) = \frac{1}{2}$ (33)

And then applying equation (9),

$I(\theta = A|x = MontyC)\ (bits) = I(A) - I(MontyC) + I(MontyC|A) = -log_2\left(\frac{1}{2}\right) + log_2\left(\frac{7}{12}\right) - log_2\left(\frac{1}{2}\right) = -\log\left(\frac{3}{7}\right) bits$ (34)

And now computing the probability associated to $I(A|MontyC)$, applying equation (5),

$P(A|MontyC) = 2^{-[I(A|MontyC)]} = 3/7$ (35)

Following the same reasoning and applying again equation (9) we get, as expected,

$P(B|MontyC)$ =4/7 (36)

Equation (35) is very surprising and totally counterintuitive: Monty by opening door C has to introduce information into the space of models, but unexpectedly the probability of door A has decreased, i.e., it seems that negative information has arrived at the mentioned door. As it will be shown below, the information contents framework does allow understanding the origin of this negative information.

In this case, when Monty opens the door C, compared to the traditional MHP in which the occurrence of car is the same in every door, does not give too much information, specifically it provides: $I(MontyC)$=$I[7/12] = -log_2\left(\frac{7}{12}\right) \approx 0{,}78 bits$. On the contrary the term correspondent to the *Information Content of the Likelihood*, $+I(MontyC|A)$, has not changed with respect to the traditional MHP, this is because independently of the prior distribution in the space of models, Monty always opens the door randomly when the prize is behind the door elected by the contestant, i.e., it always provides $1 bit$ of negative information due to the mentioned randomness that is higher than the previously commented positive information content correspondent to $P(MontyC)$, so that the net value of the two components, i.e., the *transfer information content*, is negative:

$TIC(MontyC \rightarrow A) = I(MontyC) - I(MontyC|A) = 0{,}78 - 1 = -0{,}22 bits$ (37)

Monty has generated a negative information transfer of $0{,}22 bits$ in door A when opening door C, and then, he has decreased its probability of containing the car below ½, and thus the contestant should switch to door B. As discussed later, Monty has generated a classical back action in door A that reminds to its quantum counterpart well described in (Funo, 2013).

Again, consider another variant of the traditional MHP, this is its formulation: *Monty forgets which door has the car behind it, and the contestant knows this. Monty opens either of the doors not chosen by the contestant, at random. Unluckily for the contestant, it isn't the door with the car. Should the contestant stick or switch? Answer: it doesn't matter; it's probability 1/2 either way (Rosenhouse, 2009).*

Without loss of generality we assume that Monty opens door B and again, from the formulation of the problem, we have,

$P(\theta = A) = \frac{1}{3}; P(\theta = B) = \frac{1}{3}; P(\theta = C) = 1/3$

$P(Monty_B) = P(Monty_B|A) \cdot P(A) + P(Monty_B|B) \cdot P(B) + P(Monty_B|C) \cdot P(C) = \frac{1}{2} \cdot \frac{1}{3} + 0 + 1/2 \cdot \frac{1}{3} = \frac{1}{3}$ (38)

$P(Monty_B|A) = \frac{1}{2}$ (39)

And then applying equation (9),

$I(\theta = A|x = MontyB)\ (bits) = I(A) - I(MontyB) + I(MontyB|A) = -log_2\left(\frac{1}{3}\right) + log_2\left(\frac{1}{3}\right) - log_2\left(\frac{1}{2}\right) = -\log\left(\frac{1}{2}\right) bits$ (40)

And then $P(A|MontyB)$ is given by:

$P(A|MontyB) = 2^{-[I(A|MontyB)]} = 1/2$ (41)

And thus, $P(C|MontyB) = 1/2$ (42)

This MHP variant, often known as *Forgetful Monty*, has created a lot of confusion because it is not well understood why it makes a difference that Monty opens the door without knowing a priori that there is no car behind it. The information contents approach just shown helps to understand why this variant offers different posterior information contents: again the randomness of Monty when opening the door B introduces 1bit of negative information in the system throughout the term $I(Monty_B|A)$, in this variant Monty´s behavior is always random and then it always introduces this negative information as opposed to the Traditional MHP in which Monty acts in a random way only when the contestant has elected the door with the car behind it, i.e., $1/3$ of the times.

This 1bit of negative information introduced due to the random behavior of Monty reduces the 1,58bits ($-log_2(1/3)$) of information provoked by the evidence, $I(Monty_B)$, and thus the *transfer information content* in doors A and C, due to Monty opening door B, is *only* 0,58bits. So, in this case, the evidence has been *partially informative* for both models (doors A and C):

$TIC(MontyB \rightarrow A) = I(MontyB) - I(MontyB|A) \approx 1{,}58 - 1 = 0{,}58 bits$ (43)

$TIC(MontyB \rightarrow C) = I(MontyB) - I(MontyB|C) \approx 1{,}58 - 1 = 0{,}58 bits$ (44)

## 8. QUANTUM MEASUREMENT EXPRESSED IN TERMS OF THE TRANSFER INFORMATION CONTENT

We will show that the *transfer information content* defined in equation (10) is applicable to quantum measurements. We will follow the notation in (D´Ariano 2003) and in (Terashima, 2016). (D´Ariano 2003) states that we can always regard the quantum measurement as a problem of discriminating between a set of hypotheses corresponding to an ensemble $\varepsilon(S, a)$, of states $S\{\psi\}$ distributed with a priori probability $a = a[\psi]$. After the measurement M, the a-posteriori probability $a = a[\psi]$ is given by Bayes rule (D´Ariano 2003):

$a(\psi|M) = a(\psi)P(M|\psi)/P_\varepsilon(M)$ (45)

Note that equation (10) for the transfer information content has been derived by applying the $-log_2$ function of both sides in equation (6) which is equivalent to equation (45), i.e., it is straight forward that the following equation is equivalent to equation (10):

$TIC_{QM}(M \rightarrow \psi)(bits) = I(M) - I(M|\psi)$ (46)

And from equation (9):

$I(\psi) - I(\psi|M) = TIC_{QM}(M \rightarrow \psi)$ (47)

The left side of Equation (47) is the unaveraged version of equation (2) in (D'Ariano, 2003), which is the averaged information gained from a single quantum measurement. Therefore, in equation (47), $TIC_{QM}(M \rightarrow \psi)$ represents the unaveraged information gained from a single quantum measurement, and then, the transfer information content metric defined in this work is proposed as a valid one in a Generalized Theory of Information framework (this is discussed in the next section).

## 9. SHOWING QUANTUM INDISTINGUISHABILITY BASED ON THE TRANSFER INFORMATION CONTENT (QUANTUM BACK ACTION GENERATES NEGATIVE CLASSICAL INFORMATION)

The transfer information content for quantum measurements defined in equation (46) can be applied to the problem of indistinguishability of non-orthogonal quantum states (Nielsen, 2000), and to understand the information flow to the different states.

To illustrate the information flow in a quantum measurement we will choose the indistinguishable states $|0\rangle$ and $(|0\rangle + |1\rangle)/\sqrt{2}$. When the measurement result is 1 the solution is trivial because it implies that the state must have been $(|0\rangle + |1\rangle)/\sqrt{2}$, therefore let's apply equation (46) for the case of a measurement result of 0, in this case we have:

$$\psi_1 = |0\rangle; \psi_2 = \frac{(|0\rangle + |1\rangle)}{\sqrt{2}}; M = 0;$$

$$P(\psi_1) = \frac{1}{2};\ P(\psi_2) = \frac{1}{2}$$

$$P(M=0) = P(M=0|\psi_1)\cdot P(\psi_1) + $$
$$+ P(M=0|\psi_2)\cdot P(\psi_2) = 1\cdot\frac{1}{2} + 1/2\cdot\frac{1}{2} = \frac{3}{4}$$
$$P(M=0|\psi_1) = 1$$
$$P(M=0|\psi_2) = \frac{1}{2}$$

And then, applying equation (46) to derive the bits of information transferred to each of the two quantum states (TIC) when the quantum measurement M=0, we have:

$$TIC_{QM}(M=0 \rightarrow \psi_1)(bits) = $$
$$= I(M=0) - I(M=0|\psi_1)$$
$$= -log_2\left(\frac{3}{4}\right) + log_2(1)$$
$$= 0{,}42 bits \quad (47)$$

$$TIC_{QM}(M=0 \rightarrow \psi_2)(bits) = $$
$$= I(M=0) - I(M=0|\psi_2)$$
$$= -log_2\left(\frac{3}{4}\right) + log_2\left(\frac{1}{2}\right)$$
$$= -0{,}58 bits \quad (48)$$

Now, if we express equation (9) in terms of the TIC, we can compute the a-posteriori information content of each quantum state:

$$I(\psi_1|M=0)\ (bits) = I[\psi_1](bits) - TIC_{QM}(M=0 \rightarrow \psi_1) = 1 - 0{,}42 = 0{,}58 bits \quad (49)$$

$$I(\psi_2|M=0)\ (bits) = I[\psi_2](bits) - TIC_{QM}(M=0 \rightarrow \psi_2) = 1 + 0{,}58 = 1{,}58 bits \quad (50)$$

And then, applying equation (5) to derive a posteriori probability of each quantum state:

$$P(\psi_1|M=0) = 2^{-I(\psi_1|M=0)} = 2/3 \quad (51)$$
$$P(\psi_2|M=0) = 2^{-I(\psi_2|M=0)} = 1/3 \quad (52)$$

As in the *Biased MHP*, in which Monty generates a negative information transfer of $0{,}22 bits$ in door A when opening door C, when the quantum measurement M is 0, negative bits arrive to the quantum state $\psi_2$ and its probability of occurrence decreases from 1/2 to 1/3. As in the MHP classical measurement, in the quantum case the source of the negative information corresponds to the information content of the Bayes´ likelihood, $I(M=0|\psi_2)$, which in this case is bigger than the information content of the evidence, i.e., the measurement M, and then the net information that arrives to the quantum state is negative.

# 10. TOWARDS A GENERALIZED THEORY OF INFORMATION

In (Funo, 2013), i.e., in a quantum system, the information gain is split between the information gain due to the measurement $I_{gain}$ and the information loss $I_{loss}$ due to the measurement back action. As in the classical case, in quantum systems, if the measurement back action is stronger than the acquired knowledge of the system due to the measurement, we end up with unaveraged negative information gain (Funo, 2013), (Naghiloo, 2018).

Following the previous reasoning and notation of quantum systems, in classical systems the *transfer information content* in equation (10) can be expressed as follows:

$$TIC(x \rightarrow \theta)(bits) = I(x) - I(x|\theta) = I_{gain}(evidence) - I_{loss}(Bayes\'{} likelihood) \quad (48)$$

Therefore, we have a reconciliation between classic and quantum information theories since in both cases the general *unaveraged or local mutual information (LMI)*, inside a general physical information theory, can be expressed as follows:

$$LMI_{general} = I_{gain} - I_{loss} \quad (49)$$

In both, Quantum and Classical Information Theory, the term $I_{gain}$ corresponds to the information gain due to the measurement (the evidence). With respect to the term $I_{loss}$, in both cases it represents the *inefficiency of the measurement*, and corresponds to the Information Content Associated to the Bayes´ Likelihood; and in addition, in the case of quantum systems, it represents the measurement back action well described in (Fuchs 1996), (D´Ariano 2003), (Buscemi, 2008) .

Now, if we come back to averaged information gains, it can be shown that again, equation (49) representing net information gain in a general physical theory of information, holds. Indeed, for the quantum case it is already shown in (Funo, 2013) that:

$$\langle I\rangle_{Quantum} = \langle I\rangle_{gain} - \langle I\rangle_{loss} \quad (50)$$

For classical systems, if we split equation (18) in the terms corresponding to the evidence and the ones corresponding to the Bayes´ Likelihood, we have:

$$I(\Theta, X) = H(\Theta) - H(\Theta|X) = E_x\{D_{KL}[(\Theta|X)\|\Theta]\} = E_x\{\sum_\theta P(\theta|x) \cdot TIC(x \rightarrow \theta)\} = E_x\{\sum_\theta P(\theta|x) \cdot I[P(x)]\} - E_x\{\sum_\theta P(\theta|x) \cdot I[P(x|\theta)]\} \quad (51)$$

And then, since the first term in equation (51) is always positive information, and the second one is always negative information, we have:

$$\langle I \rangle_{gain} = E_x\{\sum_\theta P(\theta|x) \cdot I[P(x)]\} \quad (52)$$

$$\langle I \rangle_{loss} = E_x\{\sum_\theta P(\theta|x) \cdot I[P(x|\theta)]\} \quad (53)$$

So, as in the quantum case, we have for the classical mutual information:

$$\langle I \rangle_{Classical} = \langle I \rangle_{gain} - \langle I \rangle_{loss} \quad (54)$$

And thus, from the similarity of (50) and (54), we have:

$$\langle I \rangle_{General} = \langle I \rangle_{gain} - \langle I \rangle_{loss} \quad (55)$$

Note that, as already discussed in this work, in the case of classical systems, $\langle I \rangle_{General}$ in equation (55) is always a non-negative quantity, as opposed to the quantum case in which $\langle I \rangle_{General}$ can be positive or negative (Funo, 2013). On the contrary, the local mutual information in equation (49) can take positive or negative values in both, classical and quantum systems.

## 11. DISCUSSION

The resolution, based on *transfer information contents*, of the previous MHP variants, and of any other similar counterintuitive problems can be generalized as follows: for a given observation, the *posterior information contents* of the models in the space are derived computing their correspondent *transfer information contents* (*and then how information transfer works is understood*), and subtracting them from their *prior information contents*. Then the model, which *posterior information* content is the least, should be chosen.

In (Lizier, 2008) it is stated that an observation can produce *negative local transfer entropy* (or equivalently *negative transfer information content*) and it is explained that in that case *"the source element is actually misleading about the state transition of the destination. It is possible for the source to be misleading in this context where other causal information sources influence the destination"*. It has been shown in this work that there is indeed other causal information that influences the destination and that this is the *Information Content Associated to the Bayes´ Likelihood* $P(x|\theta)$ that always introduces negative information which, in some cases, can be even greater than the positive information introduced by the evidence, and then, the *net information* that arrives to the mentioned destination, or the *transfer information content*, will be negative. Thus, this work finds a natural environment in the mentioned emergent field of *Local Information Dynamics* (Lizier, 2010).

It is worth to discuss in detail the expression of the Local Transfer Entropy (LTE) in equation (19) by the log-odd ratio:

$$LTE = log_2\left[\frac{P(\theta|x)}{P(\theta)}\right] \quad (56)$$

As commented in Section 6, the LTE is expressed in terms of probabilities, and thus this limits its interpretation in terms of information flow. Nevertheless, this not the only limitation of the LTE, its main drawback is that it provides the effect of data $x$ on a given parameter $\theta$, but it does not deal with the cause, or causes, of this effect. Again, the tool to that can be applied to pass from effect to causes is Bayes´ Theorem, from equation (6) it is straightforward that:

$$\frac{P(\theta|x)}{P(\theta)} = \frac{P(x|\theta)}{P(x)} \quad (57)$$

Thus, the left side of equation (57) is a function of the effects, and the right side is a function of the causes. Then, as it will be shown below, a much more useful expression of the LTE is:

$$LTE = log_2\left[\frac{P(x|\theta)}{P(x)}\right] \quad (58)$$

Now, expressing equation (58) in terms of information contents, we have:

$$LTE = log_2\left[\frac{P(x|\theta)}{P(x)}\right] = I(x) - I(x|\theta) \quad (59)$$

And from equation (10):

$$LTE = TIC \quad (60)$$

Therefore, as pointed out in Section 6, the LTE and the TIC coincide numerically; nevertheless, the former does not allow to interpret the causes of the information flow, because it is a function of probabilities and the effects, whilst the later is a function of information contents and of the causes, and therefore it provides an excellent framework for an interpretation of the information flow.

Note that, when dealing with a quantum measurement, it has been possible to express it in terms of a classical Shannon´s metric like the transfer information content instead of having used Quantum Information Theory metrics. As stated in (Terashima, 2016), the reason for this is that the uncertain information is the classical variable $a$ rather than the predefined quantum state $\psi$. In addition, the output of a quantum measurement is a probabilistic classical bit, thus it is pertinent to treat it with Classical Information Theory.

It has been shown that the information-disturbance trade-off described for quantum measurements, also takes place in classical measurements, and that, in both cases, leads to negative informative measurements. Another classical system example, in which negative unaveraged information is generated, is the Maxwell´s demon interpretation (Ito, 2016) of Sagawa-Ueda relation (Sawaga, 2010), this is the case when the demon makes a measurement error and then generates an information loss that is much bigger than the evidence.

There has been huge controversy regarding the nature of the Heisenberg principle (Benítez, 2019). Nevertheless, after an exhaustive discussion, (D'Ariano, 2003) states that the information-disturbance trade-off in a quantum measurement is a kind of Heisenberg principle, and this is inferred, for example, from the impossibility of determining the wave-function of a single system from any sequence of measurements on the same quantum system. Therefore, the fact that this information-disturbance trade-off exists as well, as shown in this paper, in classical measurements, leads to the conclusion that there exists a kind of Heisenberg principle at macroscopic scales. It is worth to mention at this point that there are generalized uncertainty principles (GUP´s) proposed in the field of quantum gravity (Maggiore, 1993), (Plato, 2016), in which the kind of Heisenberg principle at macroscopic scales described above could be included.

The novel field of thermodynamics of information (Parrondo, 2015), has shown that the Maxwell's demon is an intermediary that bridges classic information theory and thermodynamics, such as Landauer's principle and generalized second laws (Shi, 2019). In the same way, Quantum information theory (Nielsen, 2000) plays a crucial role in thermodynamics (Shi, 2019). But, unfortunately, a general physical theory of information, which reconciles classic and quantum information theories, is still missing (Parrondo, 2015). Therefore, taking into account that two toy models, the MHP and the quantum Maxwell's demon (Naghiloo, 2018), have been used to derive that both, classical and quantum unaveraged mutual information, can be negative, it would be recommendable to continue with these toy problems in order to derive the mentioned general physical theory of information, and in particular to compare the MHP with its quantum counterpart (D'Ariano, 2002). Equations (49) and (55) are proposed as a building step of this generalized theory of information.

## 12. CONCLUSIONS

Shannon´s entropy and information functions, like the KL divergence or the Mutual Information, are obtained throughout the integration in the space of models or in the space of data. Thus, all those metrics always give, as a result, the *average* of a certain concept: the average information content in the case of the entropy, the single model information gain averaged over the space of models in the case of the KL divergence or the average of the mentioned KL divergence over the space of data in the case of Mutual Information.

This work shows that, as proposed in (Hehner, 2011), considering not the averaged information content of a given space of models, but unaveraged individual information contents of each model, allows to really understand how the information, introduced by a given observation, is transferred and, as a consequence, understand counterintuitive concepts like negative information.

Expressing Bayes´ Theorem in terms of the information contents associated to the probability terms that appear in the mentioned Theorem allows computing, in the Information Theory framework, the posterior distribution in the space of models once a data is observed. It allows, as well, understanding how the information introduced in a space of models is transferred to each of them: a hypothesis, or a model, increases or decreases its probability of occurrence depending on the sign of the information that *arrives* to it.

Bayes´ Formula in terms of its information contents shows that *not one, but two information contents are generated by a single observation*: the one associated to the evidence, $I[P(x)]$ that, since it appears with the minus sign in the mentioned formula, contributes to decrease the *Posterior Information Content of the model* and to increase its probability of occurrence; and the information content of the likelihood, $I[P(x|\theta)]$ that, appears with the plus sign, and then, contributes to increase the *Posterior Information Content of the model*, and to decrease its probability of occurrence, i.e., it is the back action counterpart of classical measurements. Depending on the sign of the *transfer information content*, given by the difference $I[P(x)] - I[P(x|\theta)]$, the observation $x$ will *inform* (if the TIC>0) or *mislead* (if the TIC<0) about the single model $\theta$, i.e., as discussed later, it generates a classical back action similar to its quantum counterpart.

The Kullback-Leibler divergence can be expressed in terms of the two mentioned information contents of the evidence and the likelihood, $I[P(x)]$ and $I[P(x|\theta)]$ respectively, then it is possible to compute this metric in the Information Theory Framework; in particular the KL divergence is expressed as the average over the space of models Θ of the *transfer information content*. As in the probability framework, averaging the KL divergence over the space of data $X$, we obtain the Shannon´s Mutual Information expressed in terms of the mentioned metric defined in this work. Therefore, it is shown that it is possible to address problems, traditionally solved in the probability space, based on Shannon´s information contents.

The MHP and two of its variants have been used as an excellent example to confirm the validity of the methods proposed in this work, in particular the posterior distribution in the space of models Θ has been computed for this problem using the formulas derived in this work. In addition the mentioned MHP is an excellent framework for understanding the concept of negative information: when Monty opens a door randomly (because the contestant initially chooses the door with the car behind it), precisely due to his random behavior, he is introducing a negative information which absolute value is the information content of the likelihood, that, in some variants of the problem, may be higher than the positive information introduced by the evidence and then, the net information or the *transfer information content*, which is the one that the observer perceives, is negative. Therefore, this negative informative measurement produced by Monty is the back-action counterpart in classical systems.

Thus, the main contribution of this work is that it shows that classical information is not always positive. As its quantum counterpart, it can be negative, this is weird, and this makes more diffuse the line between classic and quantum systems. It is proposed as further work to investigate if the negativity of classical information is helpful to address the basic problems of statistical mechanics mentioned in Parrondo (2015): the complete understanding of the physical nature of information, the emergence of the macroscopic world and the subjectivity of entropy.

The fact that classic information can be negative confers it a certain degree of weirdness that reconciles it with its quantum counterpart. Indeed, the first consequence of the existence of negative informative classical measurements is a kind of Heisenberg principle at macroscopic scales. It is proposed as further work to include this result in the already proposed GUP´s in the field of quantum gravity (Maggiore, 1993), (Plato, 2016).

It is proposed, as well, as further work to build a general physical theory of information (Parrondo, 2015) based on this shared property by classical and quantum information theories. As a first step of this generalized theory of information, Equations (49) and (55) show that in both cases, classical and quantum, it is possible to model the information gained through a measurement, or the average information gained over the space of measurements, as a difference of a term, or terms in the case of averaged values, correspondent to the information gained with the evidence, and a term of negative information that accounts for the inefficiency of the mentioned measurement, which corresponds to the Bayes´ likelihood, and to the measurement back action.

Finally, again as further work, it is proposed to consider the shared property by classical and quantum information (the existence of local negative mutual information) to the problem of black hole information, raised in (Hawking, 1975), and still a topic of huge discussion (de Nova, 2019)